\documentclass[12pt]{article}
\usepackage{fullpage}
\usepackage{epsfig}
\usepackage{amsmath}
\usepackage{amssymb}
\usepackage{color}

\begin{document}


\centerline{\bf\Large A lattice study of light scalar tetraquarks}

\vspace{1cm}

\centerline{Sasa Prelovsek}

\vspace{0.1cm}

\centerline{\it Department of Physics, University of Ljubljana and Jozef Stefan Institute, Ljubljana, Slovenia}
\centerline{\tt sasa.prelovsek@ijs.si}

\vspace{0.4cm}

\centerline{and}

\vspace{0.4cm}

\centerline{Daniel Mohler}

\vspace{0.1cm}

\centerline{\it Institut f\"{u}r Physik, Universit\"{a}t Graz, 8010 Graz, Austria} 

\centerline{\tt daniel.mohler@uni-graz.at}

\vspace{1.5cm} 

\centerline{\bf Abstract}

\vspace{0.2cm}

The observed mass pattern of scalar resonances below $1$ GeV gives preference
 to the tetraquark assignment over the conventional $\bar qq$ assignment for 
these states. We present a search for tetraquarks with isospins $0,1/2,1$ 
in lattice QCD, where the isospin channels $1/2$ and $1$ have not 
been studied 
before. We determine three energy levels in each isospin channel using the variational method. The scattering states and possible tetraquark states are distinguished by considering the volume-dependence of spectral weights and by considering the time-dependence of correlators near $t\simeq T/2$.  We find no indication for light tetraquarks at our range of pion masses $344-576$ MeV.

\vspace{1.5cm}

\section{Introduction}\label{sec_intro}

In this paper we present a lattice study of  the light scalar mesons 
$\sigma$, $\kappa$, $a_0(980)$ and $f_0(980)$, whose interpretation in terms 
of $\bar qq$ or tetraquarks $\bar q\bar qqq$ is still not settled.

The resonances $a_0(980)$ and $f_0(980)$ are well established experimentally \cite{pdg08}. 
The experimental evidence  for $\sigma$ (or $f_0(600)$) has 
also become strong in recent years 
\cite{pdg08,scadron,leutwyler}. The position of the $\sigma$ pole has 
been determined  from the experimental data using 
the model independent analysis with Roy equations, 
leading to $m_\sigma=441{+16\atop -8}$ MeV and 
$\Gamma_\sigma=544{+18 \atop -215}$ MeV \cite{leutwyler}. The position of the $\kappa$ (or $K_0^*(800)$) pole has been determined using a similar analysis, leading to $m_\kappa=658\pm 13$ MeV and $\Gamma_\kappa=557\pm 24$ MeV \cite{moussallam}. Indications for $\kappa$ are also obtained from $D$, $J/\Psi$ and $\tau$ decays \cite{pdg08,kappa}, but the existence of $\kappa$ has still not been confirmed beyond doubt \cite{pdg08}. Assuming that $\kappa$ exists, these states could form a $SU(3)$ flavor nonet below 1 GeV illustrated in 
Fig. \ref{fig_spectrum}. In addition, there is a experimentally well established scalar nonet above 1 GeV as well as a few scalar glueball candidates. 

\begin{figure}[tbh!]
\begin{center}
\includegraphics[height=5cm,clip]{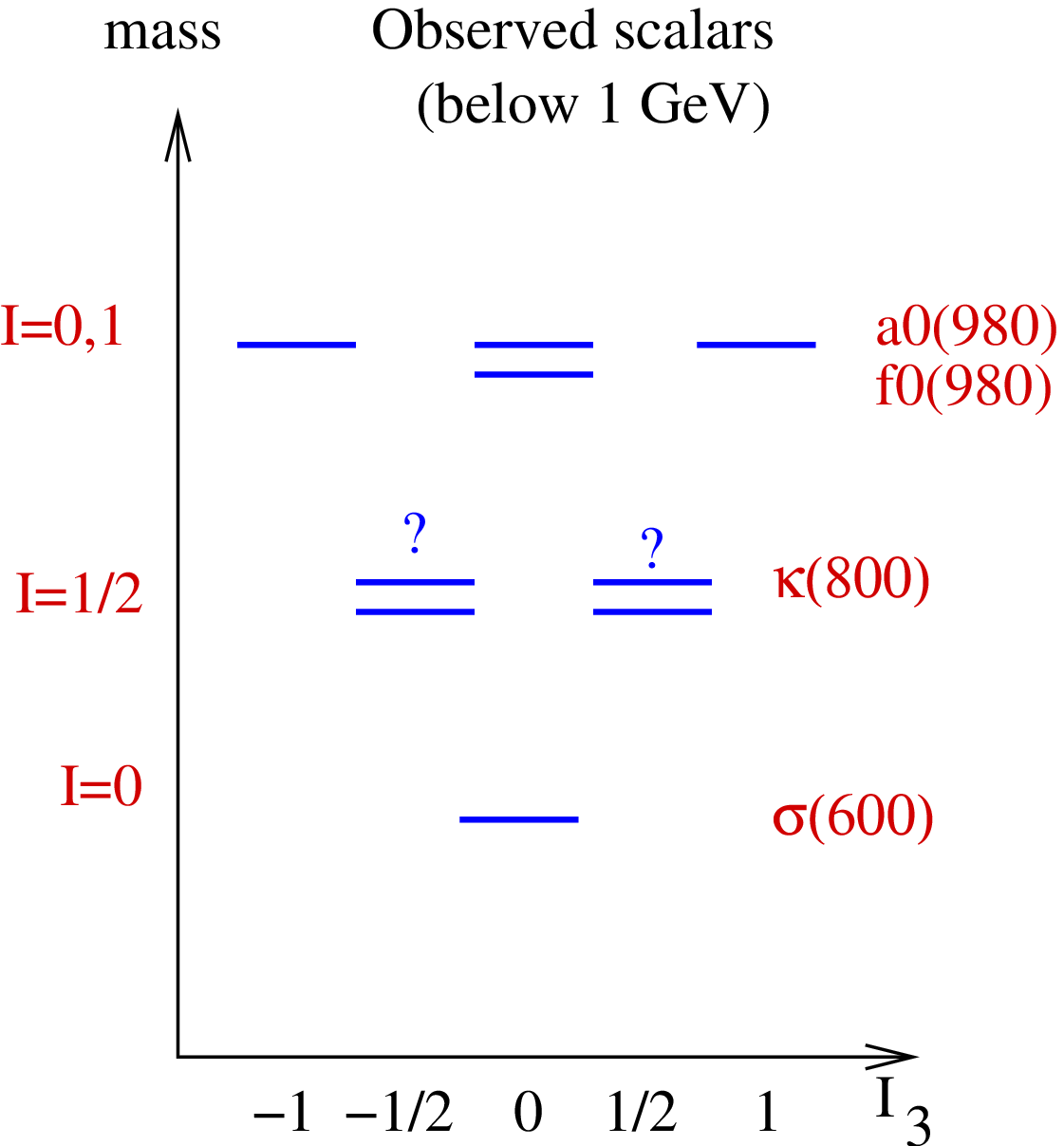}$\qquad$
\includegraphics[height=5cm,clip]{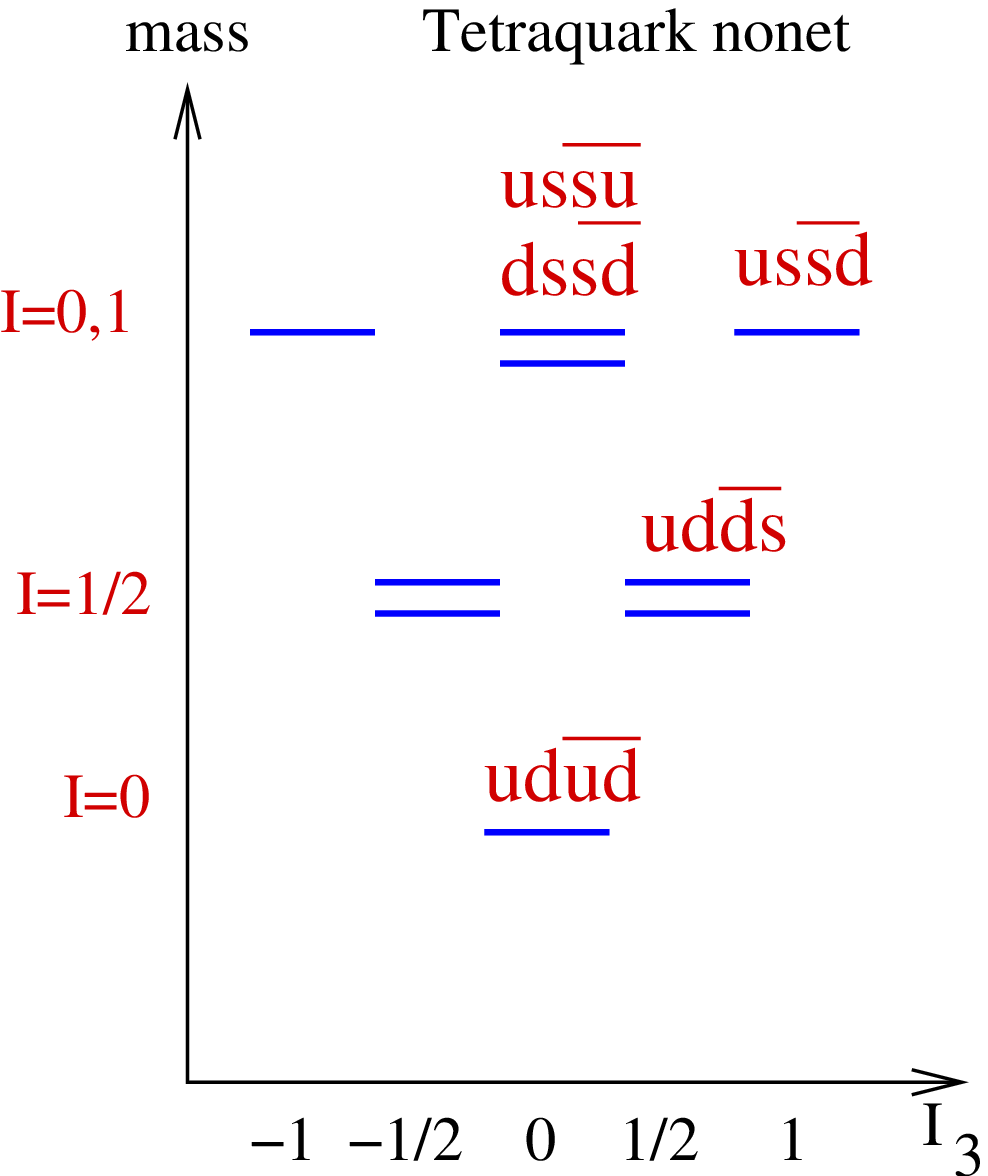}$\qquad$
\includegraphics[height=5cm,clip]{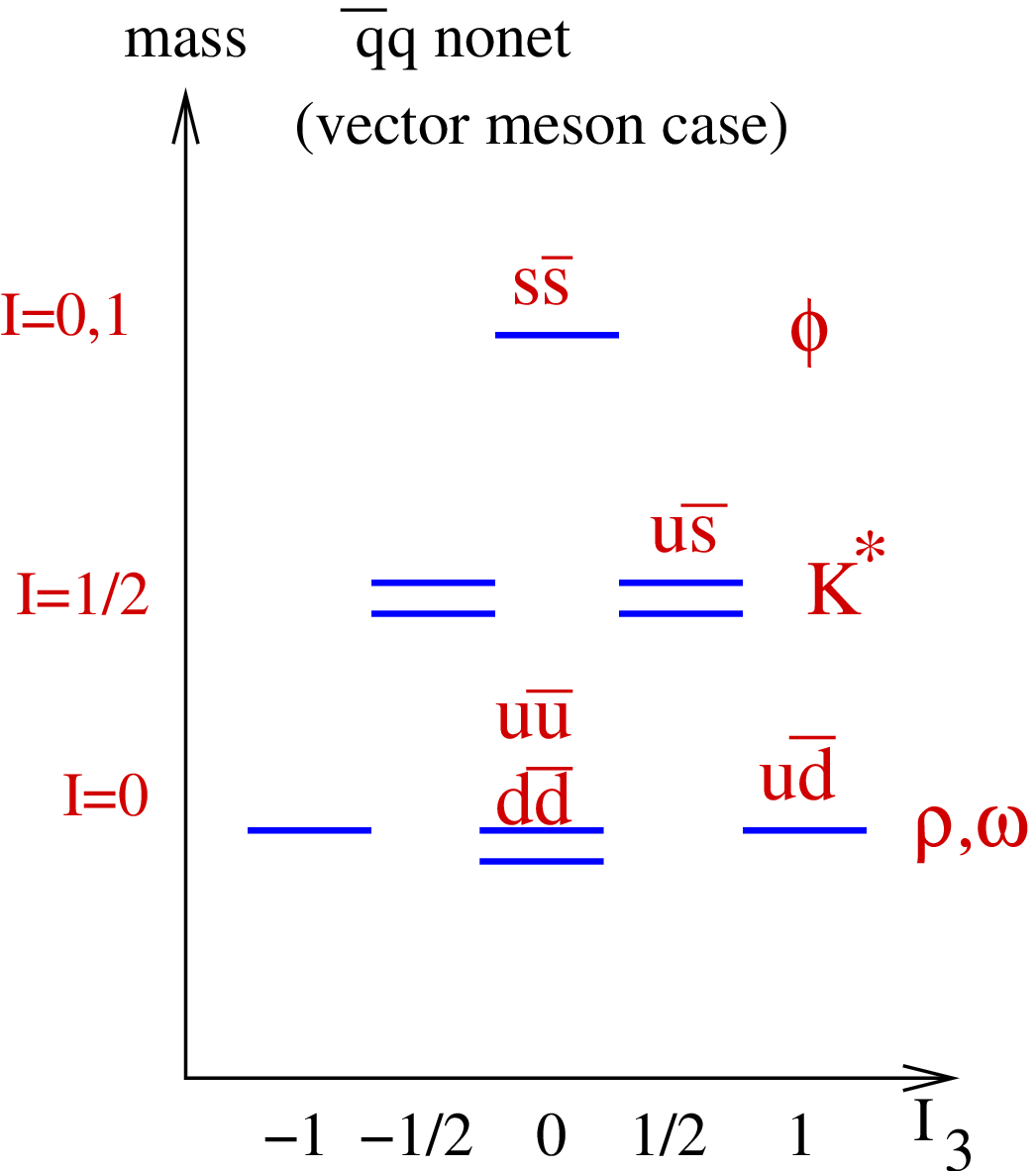}
\end{center}
\caption{ \small Schematic illustration of the observed spectrum for scalar mesons below $1$ GeV (left), together with the expected mass spectrum for the nonet of scalar tetraquarks (middle), compared with a typical $\bar qq$ spectrum (right).  }\label{fig_pattern}
\end{figure}

\begin{figure}[tbh!]
\begin{center}
\includegraphics[height=8cm,clip]{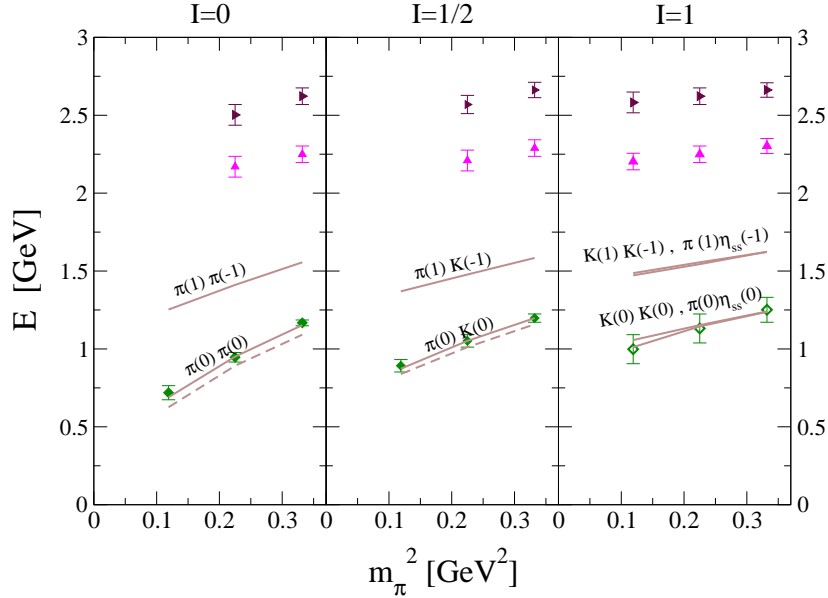}
\end{center}
\caption{ \small The symbols present the three lowest energy levels from tetraquark correlators in $I=0,1/2,1$ channels at lattice volume $16^3\times 32$. The lines give analytic energy levels for  scattering states: full lines present non-interacting energies (\ref{energies_scat}), while dashed lines take into account tree-level energy shifts (\ref{shifts}).  Appendix B explains why we don't see the energy corresponding to the scattering state $P_1(1)P_2(-1)$.}\label{fig_spectrum}
\end{figure}

The problem of identifying  all  
observed scalar mesons as conventional $\bar qq$ states has a long history. 
The proposal that the states below $1$ GeV could  actually be tetraquarks 
$\bar q\bar q qq$ dates back to Jaffe's work in 1977 \cite{jaffe77} 
and this possibility is becoming ever more vivid in recent years: 
there has been a large amount of work based on phenomenological models 
\cite{scadron,jaffe77,maiani04,hooft,tetra} with  reviews in 
 \cite{reviews}.  
   In this paper we explore the possibility 
that the nonet below 1 GeV corresponds to Jaffe-type tetraquarks 
with a diquark anti-diquark structure $[qq]_{\bar 3_f,\bar 3_c}~[\bar q\bar q]_{3_f,3_c}$. We have the scalar (or ``good'')   diquark $[qq]_{\bar 3_f,\bar 3_c}$ in mind, which is anti-triplet in flavor and  color 
and which is the most tightly bound among possible diquarks  \cite{reviews,jaffe_wilczek}. 
Combining a scalar diquark with a scalar anti-diquark $[\bar q\bar q]_{3_f,3_c}$, one gets a flavor nonet  ($\bar 3_f\times 3_f=1_f+8_f$)  of color singlet  ($\bar 3_c\times 3_c \to 1_c$)  scalar particles. The nonet is expected to be light since the (anti)diquark is tightly bound and since this is a $L=0$ state (unlike $\bar qq$ which requires $L=1$). The flavor pattern and the expected mass ordering for such a tetraquark nonet is schematically shown in Fig. \ref{fig_pattern}, along with the conventional $\bar qq$ nonet and the observed spectrum. 

Let us list just a few arguments which support a tetraquark interpretation of states below $1$ GeV, and disfavor a $\bar qq$ interpretation:
\begin{itemize}
\item Most quark models place the ground state scalar $\bar qq$ with $L=1$ above $1$ GeV, along with  well established axial and tensor $\bar qq$ mesons with $L=1$. 
\item  The observed mass ordering $m_\kappa<m_{a0(980)}$ can not be reconciled with conventional $\bar us$ and $\bar ud$ states since  $m_{\bar us}>m_{\bar ud}$ is expected due to $m_s>m_d$. This is probably the key evidence pointing to the tetraquark interpretation, where the $I=1/2$ state has the flavor structure $[ud][\bar d\bar s]$ while the $I=1$ state has the flavor structure $[us][\bar d\bar s]$. The valence $\bar ss$ pair in a $I=1$ tetraquark naturally makes this state heavier than the $I=1/2$ state and the resemblance with the observed spectrum  ($m_\kappa<m_{a0(980)}$) speaks for itself. 

Although scalar mesons above $1$ GeV are not the focus of this paper, let us mention that the observed ordering $m_{K0(1430)}<m_{a0(1450)}$ is also hard to reconcile with pure $\bar qq$ states. It has been argued that these states are mainly $\bar qq$ with a small admixture of a tetraquark state (via t'Hooft vertex), which leads to the observed mass ordering and decay pattern \cite{hooft}.  
\item The $a_0(980)$ is experimentally known to couple well with $K\bar K$, 
since this decay channel is sizable in spite of the fact that most of the phase space for $a_0\to \bar KK$ is closed. This again gives preference to  the tetraquark  over the $\bar qq$ interpretation: $[us][\bar d\bar s] \to \bar KK$ can proceed via ``cheap'' quark rearrangement  \cite{maiani04} while $u\bar d\to \bar KK$ is suppressed by the creation of $s\bar s$ pair. 
\item Recently the interesting resonances $X$, $Y$ and $Z$ containing charm quarks were discovered and some of those are serious candidates for tetraquarks. If the BELLE discovery \cite{belle_z} of charged $Z^+(4430)$  in the channel $\psi^\prime \pi^+$  is   confirmed, one has a clear indication for a four-quark state  $c\bar cu\bar d$. There have only been a few lattice studies using tetraquark interpolators aimed at these states \cite{hsieh}.  
\end{itemize}
We note that there are also other proposals for solving the scalar meson puzzle, for example  that light iso-scalars are largely glueballs \cite{minkowski} or that $a_0(980)$ and $f_0(980)$ are $K\bar K$ molecules, but these  interpretations don't naturally lead to a nonet below $1$ GeV. 

After numerous indications from experiments and from phenomenological models in favor of light scalar mesons as tetraquarks, it is very desirable to look for indications of tetraquarks in  lattice QCD. The most straight forward 
quantities for a lattice study  are the energies of the states with $J^{PC}=0^{++}$ and with quark structure $\bar q\bar qqq$ of a given flavor. The energy levels are extracted from the corresponding lattice correlation functions. The extracted energy level corresponds to the mass of a tetraquark state, if this exists. But there also unavoidably appear the energy levels corresponding to other 
$0^{++}$ states of the same flavor, for example $\pi\pi$ scattering states in case of a tetraquark $[ud][\bar u\bar d]$ with $I=0$. In order to establish the existence of a tetraquark on the lattice, one has to identify all the relevant scattering states and distinguish these scattering states from a tetraquark state. We  determine three energy levels for each isospin channel using the powerful variational method \cite{varmethod1,varmethod2,varmethod3}. The tetraquark and scattering states are distinguished by considering the volume-dependence of spectral weights and the time-dependence of correlators near $t\simeq T/2$.   

There have only been a few lattice simulation of light scalar tetraquarks  \cite{jaffe,liu, suganuma,chinese},  which are briefly reviewed in Section \ref{sec_previous}. All of them consider only the $I=0$ state, while we consider the full flavor nonet with $I=0,1/2,1$.  All previous simulations  except \cite{liu} consider only the ground state, which is most likely the $\pi\pi$ scattering state. In \cite{liu} the two excited states are extracted from a single correlator using the sequential empirical Bayes method \cite{bayes} and the first excited state appears to be  $\sigma(600)$ with a tetraquark structure. This interesting result needs  a confirmation using a different method, for example the variational method used here.

 Our main result is presented in Fig. \ref{fig_spectrum}, where the 
lowest three energy levels in each isospin channel are shown. We interpret the ground  states in all channels as scattering states, 
while the two excited states are above $2$ GeV and can not correspond to 
light tetraquarks. Therefore we find no indication for light tetraquarks 
at our $m_{u,d}$  corresponding to  $m_\pi=344-576$ MeV.
However this does not exclude the possibility of finding tetraquarks in 
 future lattice simulations with lighter quark masses or a more suitable basis of interpolators.   

\vspace{0.1cm}

Details of our simulation are presented in the next Section, while Section \ref{sec_results} provides our results. These are compared with previous lattice results in Section \ref{sec_previous}. The last section provides conclusions.

\section{Lattice simulation}\label{sec_sim}

\subsection{Correlation matrix and interpolators} 

We  calculate  tetraquark correlation functions on the lattice, where a state 
$\bar q\bar qqq$ with $0^{++}$ is created at $t=0$ and annihilated at some later time $t$. In each isospin channel $I=0,1/2,1$ we use interpolators ${\cal O}^I_{i}$ with correct quantum numbers and three different smearings $i,j=1,2,3$ that will be defined in (\ref{O_smear}) below  
\begin{equation}
\label{cor}
C_{ij}^I(t)=\langle 0| {\cal O}_i^I(t){\cal O}^{I\dagger}_j(0)|0\rangle_{\vec p =\vec 0}=\sum_{\vec x}\langle 0| {\cal O}_i^I(\vec x,t){\cal O}^{I\dagger}_j(\vec 0,0)|0\rangle~.
\end{equation} 
The correlation matrix will be used to extract energy levels of the tetraquark system with the total momentum $\vec p$ equal to zero. 

 We use theoretically well motivated diquark anti-diquark interpolators ${\cal O}^I$ \cite{jaffe77,maiani04,hooft,reviews} composed of a scalar diquark
in $\bar 3_c$ 
 \begin{equation}
[qQ]_a\equiv \epsilon _{abc} [q_b^T C\gamma_5 Q_c-Q_b^TC\gamma_5 q_c]
\end{equation}
and scalar anti-diquark in $3_c$ 
\begin{equation}
[\bar q \bar Q]_a\equiv \epsilon _{abc} [\bar q_b C\gamma_5 \bar Q^T_c-\bar Q_b C\gamma_5 \bar q_c^T]~,
\end{equation}
where $a,b,c$ are color indices. The $[ud],[us],[ds]$ have flavor 
transformations just like $\bar s,\bar d,\bar u$,  respectively, while  $[\bar u\bar d],[\bar u\bar s],[\bar d\bar s]$ transform like  $ s,d,u$. 
The combined tetraquark states $[qQ][\bar q^\prime Q^\prime]$ form a $SU(3)$ flavor nonet and are expected to be light. For concreteness we simulate particular flavor combinations 
\begin{equation}
\label{O_flavor}
{\cal O}^{I=0}=[ud][\bar u\bar d]~,\quad 
{\cal O}^{I=1/2}=[ud][\bar d\bar s]~,\quad 
{\cal O}^{I=1}=[us][\bar d\bar s]
\end{equation}
and  assume $m_u=m_d\equiv m_l$.     

 We use three different smearings of the quarks at the source ${\cal O}_j(0)$ and at the sink ${\cal O}_i(t)$ in order to extract more information from the system by means of the variational method. The three different tetraquark interpolators ${\cal O}_{i=1,2,3}$ are constructed from smeared quarks of two different widths
\begin{equation}
\label{O_smear}
{\cal O}_1^I=[q_nQ_n][\bar q_n^\prime \bar Q_n^\prime]~,\quad
{\cal O}_2^I=[q_wQ_w][\bar q_w^\prime \bar Q_w^\prime]~,\quad
{\cal O}_3^I=[q_nQ_w][\bar q_w^\prime \bar Q_n^\prime]
\end{equation}
where $q_n$ and $q_w$ refers to ``narrow'' and ``wide'' smearing, respectively. Smeared quarks are obtained from point quarks using spatially symmetric Jacobi smearing \cite{jacobi_smearing}. We use exactly the same two smearings (and same quark propagators) as applied in \cite{bgr_mesons}, which have approximately Gaussian shape and a width of a few lattice spacings (see Fig 1 of \cite{bgr_mesons}). The same smearing parameters are used on different lattice volumes.   

\subsection{Simulation details}

\begin{figure}[tbh!]
\begin{center}
\includegraphics[height=2.5cm,clip]{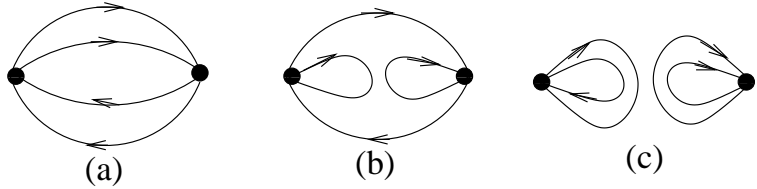}
\end{center}
\caption{ \small Quark contractions for tetraquark correlators.  }\label{fig_contractions}
\end{figure}

We use quenched gauge configurations, where the gauge fields are  generated  with the L\"uscher-Weisz gauge action \cite{luscher_weisz} and are periodic in all four directions.  All previous tetraquark simulations \cite{jaffe,liu,suganuma,chinese} were quenched as well. As pointed out by Alford and Jaffe \cite{jaffe} there is a good excuse to use the quenched approximation in this case: one is interested to look for a state with a well defined four quark content  in this pioneering era of tetraquark searches on the lattice. The quenched approximation allows for this possibility as it 
does not  mix 
$\bar qq\leftrightarrow \bar q\bar qqq\leftrightarrow vac$ via sea-quark loops. We use another approximation, which discards this mixing: in the calculation of the correlator (\ref{cor}) we omit the single (b) and double (c) annihilation diagrams among Wick contractions in Fig. \ref{fig_contractions}. 

We work on two volumes $V=L^3\times T=16^3\times 32$ and $12^3\times 24$ 
at the same lattice spacing  $a=0.148$ fm, which is determined from the Sommer parameter $r_0\simeq 0.5$ fm. 
Using different volumes is crucial as it allows the determination of volume dependence for  spectral weights.  
The quark propagators are computed from the Chirally Improved Dirac operator \cite{ci} with periodic boundary conditions in space and anti-periodic boundary conditions in time. We use $m_la=m_{u,d}a=0.02,0.04,0.06$ corresponding to $m_\pi= 344-576$ MeV. The strange quark mass is fixed to $m_sa=0.08$, which leads to a vector meson mass closest to the physical $m_{\phi}$ (see Fig. 5 of \cite{bgr_mesons}). The resulting pion and kaon masses  are given in Table \ref{tab_pseudo}. The analysis is based on 96 configuration at $16^3\times 32$ and 100 configurations at $12^3\times 24$.

\subsection{The variational method and the expected eigenstates}

Next we discuss the extraction of the energy levels from the $3\times 3$ correlation matrix  (\ref{cor}). It decomposes 
 in terms of $n=1,..N$ physical states $|n\rangle$ as follows
\begin{equation}
\label{cor_n}
C_{ij}^I(t)=\sum_n\langle 0|{\cal O}^I_i|n\rangle \langle n|{\cal O}^{I\dagger}_j|0\rangle ~e^{-E_n^I t}
\end{equation}
and the large-time correlation functions are dominated by the ground state. 
One of the physical states $|n\rangle$ is a tetraquark  if it exists. In addition there are scattering states that unavoidably couple to our interpolators (\ref{O_flavor}): $\pi\pi$ for $I=0$, $K\pi$ for $I=1/2$ and $K\bar K,~\pi\eta_{ss}$ for $I=1$ ($\eta_{ss}$ denotes a pseudoscalar meson composed of $\bar ss$). On a lattice with spatial extent $L$ and periodic boundary conditions for (anti) quarks, the energies of scattering states $P_1(\vec k)P_2(-\vec k)$ with total momentum zero come in a tower 
\begin{equation}
\label{energies_scat}
E^{P_1(j)P_2(-j)}=m_{P1}+m_{P2},\ ...\ ,\ \sqrt{m_{P1}^2+\biggl(\frac{2\pi \vec j}{L}\biggr)^2}+\sqrt{m_{P2}^2+\biggl(\frac{2\pi \vec j}{L}\biggr)^2},...
\end{equation}
for integer $\vec j$ in assuming non-interacting pseudoscalars. The lowest few energy levels are well separated at the volumes we are using. In the interacting theory, the energy levels  in a finite box are shifted by L\"uscher shifts \cite{luscher}. These are expressed in terms of infinite-volume scattering lenghts and are given by \cite{orginos_scat} 
\begin{equation}
\label{shifts}
dE_{\pi\pi,I=0}^{tree}=-7/(4f_\pi^2 L^3)\qquad dE_{\pi K,I=1/2}^{tree}=-1/(f_\pi^2 L^3)
\end{equation}
at tree-level\footnote{L{\"u}scher \cite{luscher} provides shifts in full QCD. The tree-level shifts in quenched QCD agree with full QCD. The quenched shifts beyond tree-level have only been derived for the case of $\pi\pi$ scattering \cite{bernard_golterman}.}.

In order to show the existence or absence of tetraquarks on the lattice, one has to establish energy-levels for a tetraquark as well as for the few lowest  scattering states. Since fitting the sub-leading exponentials in (\ref{cor_n}) is very unstable, we use the variational method. We therefore compute the eigenvalues and eigenvectors which diagonalize the $3\times 3$ hermitian correlation matrix\footnote{All our correlators $C_{ij}(t)$ are symmetric with respect to $t\leftrightarrow T-t$ and we fold them  as $\tfrac{1}{2}[C_{ij}(t)+C_{ij}(T-t)]$ before computing the eigenvalues.}
\begin{equation}
\label{var}
C(t)\vec v_n(t)=\lambda_n(t)\vec v_n(t)~
\end{equation}
at each time slice. The eigenvalues are dominated by
 a single physical state \cite{varmethod2}
\begin{equation}
\label{lambda_n}
\lambda_n(t)=w_n e^{-E_nt} ~(1+{\cal O}(e^{-\Delta_n t}))
\end{equation}
which can be easily understood if there were only $N=3$ physical states. 
The corrections are given by the energy difference to the nearest energy level $\Delta_n$.  The eigenvectors $\vec v_n(t)$ are orthogonal and represent the components of physical states in terms of the chosen three-dimensional variational basis (\ref{O_smear}). We also have determined the energy levels using the 
generalized eigenvalue problem $C(t_0)^{-1/2}C(t)C(t_0)^{-1/2}\vec v_n^\prime(t)=\lambda_n^\prime(t)\vec v_n^\prime(t)$ for various values of $t_0$ and the extracted energies agree with the results from the standard eigenvalue problem (\ref{var}) withing the errors. In the following we only present results based on the standard eigenvalue problem, where the determination of $w$ as a function of $L$ is more direct\footnote{In the generalized problem   $C(t_0)$ defines the normalization and $\lambda^\prime_n(t_0)=1$ by construction.}. 

All eigenvalues are fitted using uncorrelated fits, which explains a rather small $\chi^2/d.o.f$ in the Tables. All quoted errors are statistical and are determined using the jackknife method.

\section{Results}\label{sec_results}

\subsection{I=0}

{\bf Energies}

\vspace{0.1cm}

In this channel we need to identify the $\pi\pi$ scattering states and find out if there is an additional candidate for  $\sigma$ with the content $[ud][\bar u\bar d]$.  

\begin{figure}[tbh!]
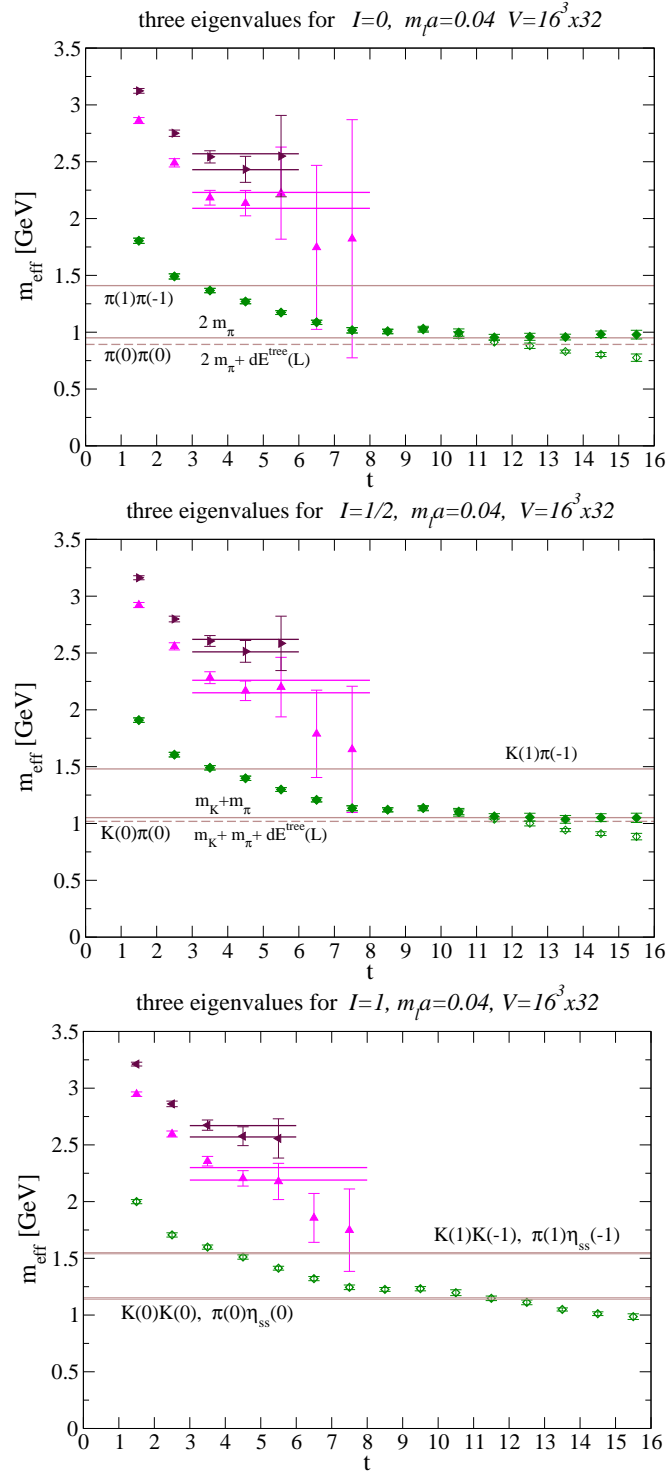

\begin{center}
\includegraphics[height=6.5cm,clip]{fig/eff_scalar_I_zero_4q_ml_0.04_mh_0.08_phy_16x32.eps}

\includegraphics[height=6.5cm,clip]{fig/eff_scalar_I_half_4q_ml_0.04_mh_0.08_phy_16x32.eps}

\includegraphics[height=6.5cm,clip]{fig/eff_scalar_I_one__4q_ml_0.04_mh_0.08_phy_16x32.eps}
\end{center}
\caption{ \small Effective masses for the three eigenvalues $\lambda_{0,1,2}(t)$ in $I=0,1/2,1$ channels for $m_la=0.04$ and $V=16^3\times 32$. The full diamonds were obtained using $m_{eff}$ defined in (\ref{meff_bc},\ref{meff_bc2}), while  others were obtained using conventional cosh-like definition (\ref{meff}). The lines give energy levels for  scattering states: full lines present non-interacting energies (\ref{energies_scat}), while dashed lines take into account tree-level energy shifts (\ref{shifts}).}\label{fig_eigenvalues}
\end{figure}

\clearpage

\begin{figure}[tbh!]
\begin{center}
\includegraphics[height=7.5cm,clip]{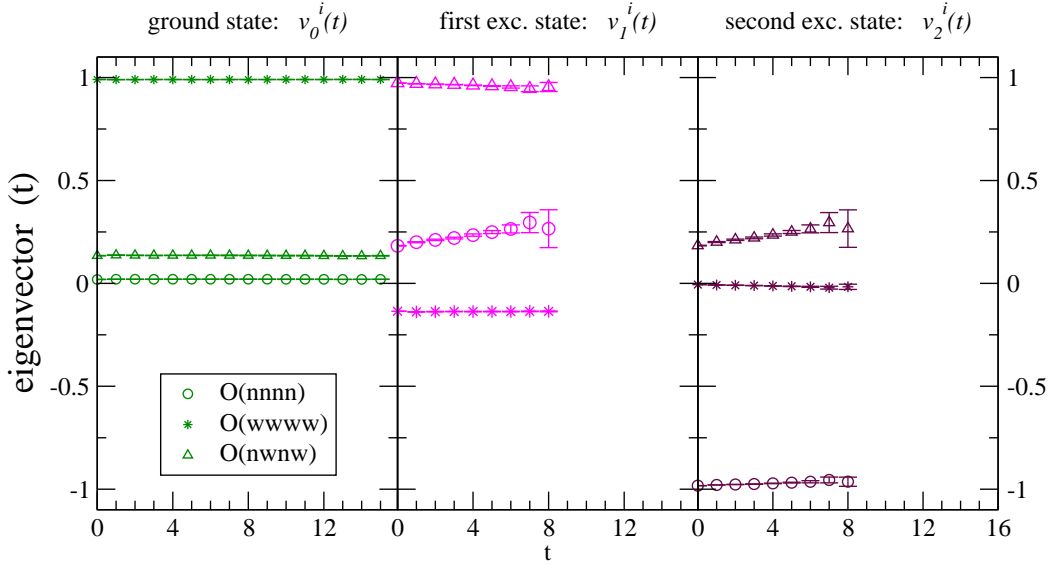}
\end{center}
\caption{ \small Eigenvector components $Re(v^i_n)$ as a function of $t$ for three eigenstates $n=0,1,2$ in terms of the basis ${\cal O}_i$ (\ref{O_smear}). We find  $Im(v^i_n)\simeq 0$.  The figure shows the case $I=0$, $m_l a=0.04$ and $V=16^3\times 32$ and others are very similar: eigenvectors show almost no isospin, quark mass or volume dependence. }\label{fig_I_zero_eigenvectors}
\end{figure}

The three lowest energy levels are given by 
the effective masses for the three eigenvalues $\lambda_n(t)$ at large $t$. 
They  are plotted in Fig. \ref{fig_eigenvalues}a for the case of $m_l=0.04$ and $V=16^3\times 32$. The corresponding eigenvectors, given in Fig. \ref{fig_I_zero_eigenvectors}, illustrate the decomposition of the three eigenstates in terms of our basis (\ref{O_smear}): $|n\rangle=v_n^i |{\cal O}_i\rangle$. The energy of the ground state is close to $2m_\pi$, indicating that the ground state is  a $\pi(0)\pi(0)$ scattering state. Below we will present two other arguments which support this interpretation of the ground state.  The most important finding is that there is a large gap above the ground state: the first and the second excited states appear only at energies above $2$ GeV. Whatever the nature of these two excited states is, they are much to heavy to correspond to $\sigma(600)$, which is the light tetraquark candidate in this channel.  
The two excited states may correspond to  $\pi(\vec k)\pi(-\vec k)$  with higher $\vec k$ or to some other energetic state. We refrain from identifying them with certain physical objects since  such energetic states are not the focus of our present study. 

A very similar conclusion applies for the other light quark masses and for the other volume. All the ground state energy levels are close to $2m_\pi$, while the excited energy levels $E_{1,2}$ are above $2$ GeV, as shown in Fig. \ref{fig_spectrum} and Table \ref{tab_I_zero} \footnote{The fitting forms for the extraction of energies are presented in the next subsection and in Appendix A.}.
 Apart from $\pi(0)\pi(0)$, we find no state with energy close to $m_\sigma$, so we find no indication for a light tetraquark in the $I=0$ channel. Since all our pion masses are just above $300$ MeV, this is not in conflict with the simulation of the Kentucky group \cite{liu}, which finds indication for a tetraquark with mass $\sim 550$ MeV at $m_\pi=180-300$ MeV (but not above that). 

Looking at the spectra in Figs. \ref{fig_eigenvalues} and \ref{fig_spectrum}
the question araises why  there is no 
state close to the energy $2\sqrt{m_\pi^2+(2\pi/L)^2}$ of a $\pi(1)\pi(-1)$ state. In Appendix B we argue that the tower of few lowest scattering states $\pi(\vec k)\pi(-\vec k)$ contributes to the lowest eigenvalue, which explains why its effective mass is not flat at intermediate $t$. This is due to the fact that  our basis (\ref{O_smear}) does 
not disentangle the few lowest scattering states into separate eigenvalues. They  would contribute to separate eigenvalues by using a larger basis or using interpolators with a definite momentum projection for each pion.\\

{\bf Time-dependence of the eigenvalues}

\vspace{0.1cm}

We were surprised to find that the ground state eigenvalues do not have a conventional time-dependence $w[e^{-E t}+e^{-E (T-t)}]$ near $t\simeq T/2$. This was first noticed by looking at the cosh-type effective  mass (\ref{meff}), which is decreasing near $t\simeq T/2$  (see empty symbols in Figs. \ref{fig_eigenvalues} and \ref{fig_corrected}). In Appendix A we derive that the eigenvalues for the $\pi\pi$ state receive a constant in addition to the conventional term $w[e^{-E t}+e^{-E (T-t)}]$ (\ref{lambda_bc}). Our data for $\lambda_0^{I=0}(t)$ agrees well with this analytic expectation  and the effective mass (\ref{meff_bc}) which takes this into account   
has a plateau near $t\simeq T/2$ (see full symbols in Figs. \ref{fig_eigenvalues} and \ref{fig_corrected}). The non-conventional time-dependence of the ground state is the second indication that this is a scattering state. The values of $E_0$ and $w_0$ in Table \ref{tab_I_zero} are obtained from the three-parameter fit of $\lambda_0(t)$  (\ref{lambda_bc}). The presence of an additional constant increases the uncertainty of the resulting $E_0$ and this is the main reason that we refrain from studying  the $\pi\pi$ energy shifts. 

All  excited states are obtained from the conventional fit to $\lambda_{1,2}(t)=w_{1,2}[e^{-E_{1,2}t}+e^{-E_{1,2}(T-t)}]$.  \\

{\bf Spectral weights}

\vspace{0.1cm} 

Let us describe a method, which considers $w(L)$ in order to 
allow the distinction between 
 the one-particle (tetraquark) and the 
two-particle (scattering) states on the lattice.

\begin{figure}[tbh]
\begin{center}
\includegraphics[height=2cm,clip]{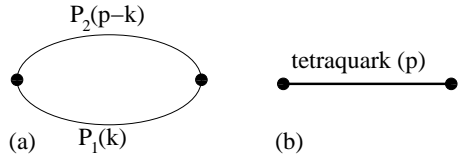}
\end{center}
\caption{ \small Diagrams relevant for deriving the volume dependence of spectral weight for two-particle (a) and one-particle (b) state contributions.}\label{fig_diag_w}
\end{figure}

The $L$ dependence of the spectral weight $w$ for a scattering state $P_1P_2$ has been derived in coordinate space in Section III E of \cite{kentucky_penta}. We re-derive it in momentum space by  evaluating the contribution in Fig. \ref{fig_diag_w}a for finite $L$, infinite $T$ and $\vec p=\vec 0$ 
\begin{align}
\label{derive_w}
\langle{\cal O}_i(t)|P_1P_2\rangle\langle P_1P_2|O_i^\dagger (0) \rangle & \propto \int dp_4 ~ e^{ip_4 t} \int\frac{d\vec k}{(2\pi )^3} \frac{dk_4}{2\pi }\frac{|\langle{\cal O}_i|P_1(\vec k)P_2(-\vec k)\rangle|^2 }{[k_4^2+\vec k^2+m_1^2][(p_4-k_4)^2+\vec k^2+m_2^2]} \nonumber \\
&\propto \frac{1}{L^3}\sum_{\vec j} |\langle{\cal O}_i|P_1(\vec j)P_2(-\vec j)\rangle|^2 \frac{e^{-\left (E^{\vec j}_1+E^{\vec j}_2\right )t}}{E^{\vec j}_1~E^{\vec j}_2}~.
 \end{align}
The leading volume dependence of the coefficient in front of the exponent  (called the spectral weight $w$) for a given physical state $\vec j$ is 
\begin{equation}
\label{w}
w^{P_1(\vec j)P_2(\vec j)}\propto 1/L^3
\end{equation}
  and it comes from  $dk_j=2\pi/L$. The energies $E^{\vec j\not =\vec 0}=\sqrt{m^2+(2\pi \vec j/L)^2}$ in (\ref{derive_w}) 
are only mildly dependent on $L$.    
The couplings $\langle{\cal O}_i|P_1(\vec j)P_2(-\vec j)\rangle$ at given $\vec j$ are  volume independent when the size and the shape of interpolators do not depend on $L$ (which is true for our interpolators) and when the coupling is non-derivative. In the case of derivative coupling  \cite{hooft}, which is proportional to $k_{P_1}^\mu k_{P_2\mu}=\sqrt{m_1^2+(2\pi \vec j/L)^2}\sqrt{m_2^2+(2\pi \vec j/L)^2}+(2\pi \vec j/L)^2$, there is a mild dependence on $L$ for $\vec j\not =\vec 0$, but this dependence is sub-leading with respect to the $1/L^3$ dependence  (\ref{w}).

 The contribution of a one-particle  (tetraquark) state in Fig. \ref{fig_diag_w}b gives 
\begin{equation}
\langle{\cal O}_i(t)|T\rangle\langle T|O_i^\dagger (0)  \rangle  \propto  \int dp_4 ~e^{ip_4 t}~ \frac{1}{p_4^2+m^2} \propto \frac{e^{-m  t}}{m}
\end{equation}
and the spectral weight is expected to be almost independent of the volume.

\begin{figure}[tbh]
\begin{center}
\includegraphics[height=7.5cm,clip]{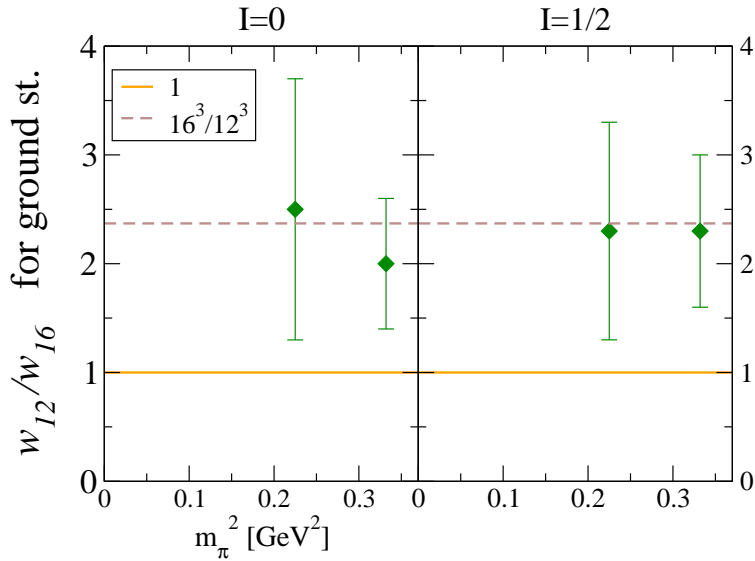}
\end{center}
\caption{ \small The ratio of spectral weights $w_0(L=12)/w_0(L=16)$ 
for $I=0,1/2$ as computed from the ground state eigenvalues for two volumes $L^3$.
}\label{fig_w}
\end{figure} 

Our spectral weights $w$   
are presented for both volumes  in Table \ref{tab_I_zero}.  The ratio 
 $w_0(L=12)/w_0(L=16)$ for the ground state is plotted  in Fig. \ref{fig_w}, 
where the individual errors are summed  in quadrature
 $\Delta(w_{12}/w_{16})=\sqrt{|\Delta w_{12}/w_{16}|^2+|\Delta w_{16}~w_{12}/w_{16}^2|^2}$. 
The ratio is consistent with $16^3/12^3$ within large error 
bars and its $w\propto 1/L^3$ dependence is the third indication that our ground state is a $\pi\pi$ scattering state. 

\subsection{$I=1/2$}

The conclusions regarding $I=1/2$ are very similar to $I=0$. 

The ground state energy in is close to $m_\pi+m_K$ (see Figs. \ref{fig_eigenvalues} and \ref{fig_spectrum}), which indicates it is a $\pi(0) K(0)$ state. The corresponding eigenvalue has a non-conventional time dependence near 
$t\simeq T/2$, which agrees with the analytic expectation (\ref{lambda_bc2}) 
 for a $\pi(0) K(0)$ state. The cosh-type effective mass does not have a plateau near $t\simeq T/2$,  while the  corrected effective mass (\ref{meff_bc2}) has a plateau (see Fig. \ref{fig_corrected}). The three parameter fit of $\lambda_0(t)$ (\ref{lambda_bc2}) gives $E_0$ and $w_0$ in Table \ref{tab_I_half}. The volume dependence of the spectral weight for the ground state $w_0(L=12)/w_0(L=16)$ is consistent with $16^3/12^3$ (see Fig. \ref{fig_w}) which is another indication for a $K\pi$ scattering state. 

The energies of the first and the second excited states in Figs. \ref{fig_eigenvalues} and \ref{fig_spectrum} are above $2$ GeV.  So we do not find an indication of a tetraquark with mass close to the mass of $\kappa(800)$. 

\subsection{$I=1$}

There are two scattering states  $K\bar K$ and $\pi\eta_{ss}$ with similar energies at $I=1$, which makes this channel more challenging. The major features of the results are the same as for $I=0,1/2$: the ground state energy in Figs.  \ref{fig_spectrum} and \ref{fig_eigenvalues} is close to $2m_K$ and $m_\pi+m_{\eta_{ss}}=m_\pi+m_{\pi_{ss}}$, while the excited states are above $2$ GeV. Again, we do not find an indication for a tetraquark with mass close to the mass of $a_0(980)$.

However, in this channel our conclusion is not so firm since we were not able to analyze $\lambda_0(t)$  using a two-state fit with $K\bar K$ (\ref{lambda_bc}) and $\pi \eta_{ss}$ (\ref{lambda_bc2}). Such a fit is not stable with our data, especially since the two states are close in energy and since there are additional terms due to finite $T$ in (\ref{lambda_bc},\ref{lambda_bc2}). 
We present only approximate ranges\footnote{The naive cosh-type effective mass (\ref{meff}) for $V=16^3\times 32$ has a brief plateau at $t=8-10$ and a sizable fall-off for $t=10-16$. We give approximate  
 ranges of ground state energies as $m_{eff}^{9.5}-m_{eff}^{13.5}$ in Table \ref{tab_I_one} and Fig. \ref{fig_spectrum}, which should serve only as guidance.} for the ground state energies  in Table \ref{tab_I_one} and Fig. \ref{fig_spectrum}, while we are unable to  present reliable values for the corresponding spectral weights.

Due to the presence of two towers of scattering states in addition to a possible tetraquark, the $I=1$ channel remains a challenge with the methods available at present, even for future simulations.  

 \section{Previous lattice simulations of light tetraquarks }\label{sec_previous}

Finally we compare our results with the results of previous simulations 
 \cite{jaffe,liu,suganuma,chinese}. Like ours, all these simulations are 
quenched and omit the contributions of annihilation diagrams (b), (c)  
in Fig \ref{fig_contractions}. In contrast to the present work, they all 
consider only $I=0$ \footnote{The authors of \cite{jaffe,chinese} also consider the $I=2$ channel, but this does not belong to the nonet considered here.} and only the ground state (except for \cite{liu}). 

\vspace{0.1cm}

Alford and Jaffe \cite{jaffe} computed the ground state energy $E_0$  for $\pi\pi$ interpolators at fixed $m_\pi\simeq 800$ MeV and various  $L$.  They find that the energy shift $E_0-2m_\pi$ does not completely agree with the (full ChPT) analytic prediction for $\pi\pi$ scattering at finite volume. They interpret this as a possible indication for a tetraquark. 

The Kentucky group \cite{liu} computed the $\pi\pi$ correlator using overlap quarks for $m_\pi=182-760$ MeV and two volumes $L^3$. 
Using the sequential empirical Bayes method \cite{bayes} 
they were able to extract 
three states from a single correlator. The ground state with $E_0\simeq 2m_\pi+dE(L)$ and $w_0(L=12)/w_0(L=16)\simeq 16^3/12^3$  is interpreted as $\pi(0)\pi(0)$. The first excited state with $E_1\simeq 600$ MeV and $w_1(L=12)/w_1(L=16)\simeq 1$ gives indication for $\sigma$ with a tetraquark structure. They find this state only for $m_\pi\leq 300$ MeV, so their results are not in contradiction with our results, which give no evidence for a tetraquark at $m_\pi\geq 300$ MeV.  Their second excited state has $E_2\simeq E^{\pi(1)\pi(-1)}$ and is interpreted as $\pi(1)\pi(-1)$. This is one of the few available lattice studies which was able to separate the one and two-particle contributions and gives impressive evidence for $\sigma$ as a tetraquark. However, it relies on the ability of the  sequential method to extract three states from a single correlator and therefore needs confirmation using other methods. 

The Japanese group \cite{suganuma} determined the ground state energy for a diquark-antidiquark interpolator $[ud][\bar u\bar d]$ at $m_{u,d}=[m_s,2m_s]$. They find that the extracted energy is close to the $\pi\pi$ energy  in the case of periodic as well as in the case of hybrid boundary conditions. They conclude the ground state is not a tetraquark state.    

The authors of \cite{chinese} use the variational method and extract only the energy of the ground state, which is close to $2m_\pi$ for their range of pion masses.  

\section{Conclusions and outlook} \label{sec_conclucions}

The question whether the observed scalar resonances $a_0(980)$, $f_0(980)$, $\sigma(600)$ and $\kappa(800)$ correspond to conventional $\bar qq$ or tetraquark nonet is still not settled. The observed mass pattern gives preference to the later. 

We presented a search for possible light scalar tetraquarks by means 
of a quenched lattice simulation.  
In each isospin channel we extracted the three lowest states from a $3\times 3$ 
correlation matrix with diquark-antidiquark interpolators. 
The energy levels are shown in  
Fig. \ref{fig_spectrum}, which is the main result of our paper. 
 The ground states are found to be consistent with the scattering 
states.  The first and second excited states have energies above $2$ GeV, so they can not correspond to resonances below $1$ GeV. 

In conclusion, we  find no indication for light tetraquarks at our range of
 pion masses $344-576$ MeV. However, one should not give up hopes for finding these interesting objects on the lattice. Indeed, our simulation does not exclude the possibility of finding tetraquarks for lighter $m_{u,d}$ or for a larger (different) interpolator basis. A stimulating lattice indication for $\sigma$ as a tetraquark state at $m_\pi=182-300$  MeV has already been presented in \cite{liu}. 

The present and past pioneering quenched tetraquark simulations, which discard annihilation diagrams, provide valuable information on the states with a definite quark assignment. The final conclusions  will have to await dynamical simulations incorporating both annihilation quark diagrams and the $\bar q\bar qqq\leftrightarrow \bar qq \leftrightarrow vac$ mixing. 

\vspace{1cm}

{\bf Acknowledgments}

\vspace{0.2cm}

We would like to thank the BGR Collaboration for providing the gauge configurations and quark propagators used for this project. Special thanks goes to  
C. Lang and C. Gattringer for numerous valuable discussions. We thank M. Savage and W. Detmold for pointing us to their paper which considers the interesting artifact described in our Appendix A. We are also grateful for valuable discussion with Keh-Fei Liu, T. Draper, N. Mathur, S. Fajfer and T. Burch. This work is supported in part by European RTN network, contract number MRTN-CT-035482 (FLAVIAnet).  D.M. is supported by the DK W1203-N08 of the "Fonds zur F\"orderung wissenschaflicher Forschung in \"Osterreich''.

\appendix
\section{Time dependence of correlators for scattering states at finite $T$}

\begin{figure}[tbh!]
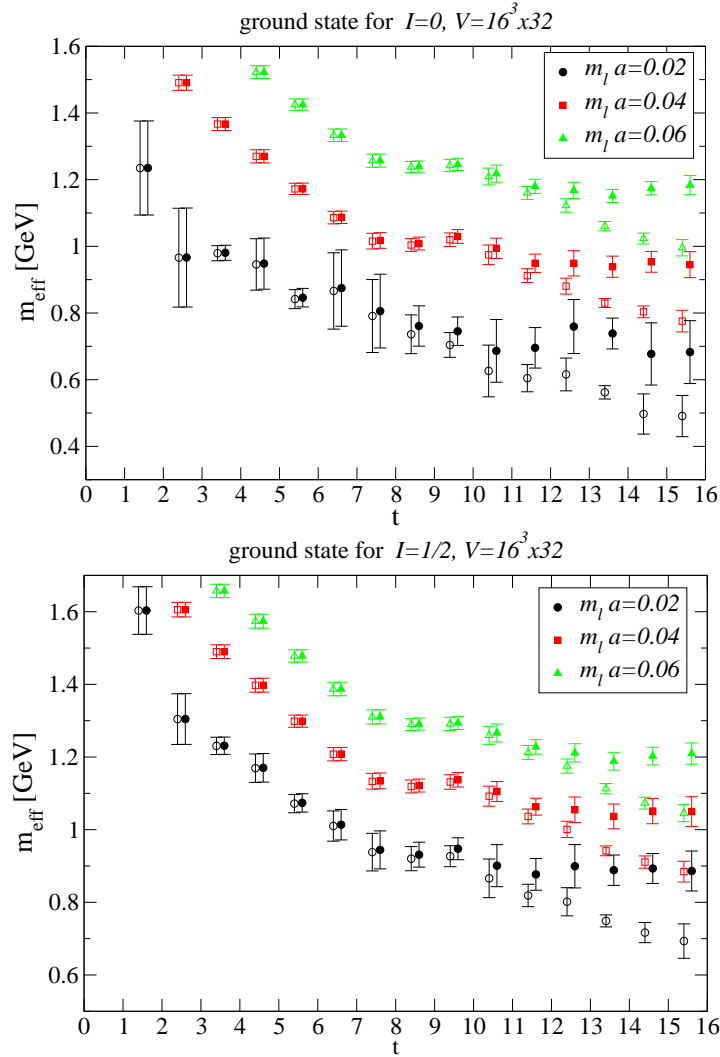

\begin{center}
\includegraphics[height=7cm,clip]{fig/eff_1_scalar_I_zero_phy_bc_16x32.eps}
\includegraphics[height=7cm,clip]{fig/eff_1_scalar_I_half_phy_bc2_16x32.eps}
\end{center}
\caption{ \small Effective masses for the $I=0,1/2$ ground states at various $m_la$ and $V=16^3\times 32$. Empty symbols are obtained using ``naive'' cosh-like definition  for $m_{eff}$ (\ref{meff}), while full symbols take into account the correct form: (\ref{meff_bc})  for $\pi\pi$ scattering and (\ref{meff_bc2}) for $K\pi$ scattering. }\label{fig_corrected}
\end{figure}

In this Appendix we discuss an interesting observation, related to scattering states on the lattice with finite time extent $T$. We noticed an unconventional behavior by looking at the cosh-type effective mass
\begin{equation}
\label{meff}
\frac{\lambda(t)}{\lambda(t+1)}=\frac{e^{-m_{eff}^{t+1/2}~t}+e^{-m_{eff}^{t+1/2}~(T-t)}}{e^{-m_{eff}^{t+1/2}~(t+1)}+e^{-m_{eff}^{t+1/2}~(T-t-1)}}
\end{equation}
for the ground state, which keeps falling even at large $t$ close to $T/2$ (see empty symbols in Figs. \ref{fig_eigenvalues} and \ref{fig_corrected}). This indicates that $\lambda_0(t)$ does not have the conventional time-dependence $e^{-E_0t}+e^{-E_0(T-t)}$ at large $t$. 

We believe this is due to the fact that the ground state is a scattering state $P_1P_2$. Let us derive the time dependence of a correlator $C(t)$  which creates and annihilates a state by the interpolator ${\cal O}$. We consider the case where the lowest physical state coupling to ${\cal O}$ is the scattering state $P_1P_2$. An example is  $P_1P_2=\pi^+K^0$ with ${\cal O}=[ud][\bar d\bar s]$ or ${\cal O}=\pi^+K^0$, which is relevant to our correlators with $I=1/2$. We start from the basic definition of $C(t)$ \cite{detar}
\begin{eqnarray}
C(t)&=&\frac{1}{Z}Tr\bigl[e^{-HT}{\cal O}(t){\cal O}^\dagger (0)\bigr]= 
\frac{1}{Z}Tr\bigl[e^{-H(T-t)}{\cal O}e^{-Ht}{\cal O}^\dagger\bigr]\nonumber\\
&\equiv&\frac{1}{Z}\sum_m \langle m|e^{-H(T-t)}{\cal O}e^{-Ht}{\cal O}^\dagger|m\rangle
= \frac{1}{Z}\sum_{m,n} \langle m|e^{-H(T-t)}{\cal O}|n\rangle\langle n|e^{-Ht}{\cal O}^\dagger|m\rangle~.
\end{eqnarray}
When $t$ and $T-t$ are large, only the states $n,m=0,P_1P_2,P_1,P_2$ 
contribute and the non-vanishing terms in this limit are
\begin{eqnarray}
C(t)&= &\frac{1}{Z}\biggl(\langle 0|e^{-H(T-t)}{\cal O}|P_1P_2\rangle
                       \langle P_1P_2|e^{-Ht}{\cal O}^\dagger|0\rangle 
 +\langle P_1^\dagger P_2^\dagger|e^{-H(T-t)}{\cal O}|0\rangle
\langle 0|e^{-Ht}{\cal O}^\dagger|P_1^\dagger P_2^\dagger\rangle\nonumber\\
&~&\ \ +\ \langle P_1^\dagger|e^{-H(T-t)}{\cal O}|P_2\rangle
 \langle P_2|e^{-Ht}{\cal O}^\dagger|P_1^\dagger\rangle
+\langle P_2^\dagger|e^{-H(T-t)}{\cal O}|P_1\rangle
\langle P_1|e^{-Ht}{\cal O}^\dagger|P_2^\dagger\rangle \biggr)\nonumber\\
&=&\frac{1}{Z}\biggl(|\langle 0|{\cal O}|P_1P_2\rangle|^2 e^{-E^{P_1P_2}~t}
+|\langle P_1^\dagger P_2^\dagger|{\cal O}|0\rangle|^2 e^{-E^{P_1P_2}(T-t)}\nonumber\\
&~& \ \  +\ |\langle P_1^\dagger|{\cal O}|P_2\rangle|^2 e^{-E_{P_1}(T-t)}e^{-E_{P_2}t}
 +|\langle P_2^\dagger|{\cal O}|P_1\rangle|^2 e^{-E_{P_2}(T-t)}e^{-E_{P_1} t}\biggl)~,
\end{eqnarray}
where $H|0\rangle=0$. In first term $P_1$ and $P_2$ propagate forward in time, in the second they both propagate backward, while in the third and fourth term one propagates forward and the other backward. 
For the case of our ground  state $P_1(\vec 0)P_2(\vec 0)$ with the anti-periodic propagators in time this gives the appropriate form to fit the eigenvalue
 \begin{equation}
\label{c_bc}
\lambda_0^{P_1(0)P_2(0)}(t)=w [ e^{-E_0 t}+e^{-E_0 (T-t)}]+ A [e^{-m_{P1} t}e^{-m_{P2} (T-t)} + e^{-m_{P2} t}e^{-m_{P1} (T-t)}] 
\end{equation} 
with $E_0\simeq m_{P1}+m_{P2}$. A similar relation was provided without derivation in the Appendix of \cite{savage}. 
 
Therefore we fit the eigenvalue for the ground state $\pi(0)\pi(0)$ to 
\begin{equation}
\label{lambda_bc}
\lambda_0^{\pi\pi}(t)=w_0 [ e^{-E_0 t}+e^{-E_0 (T-t)}]+ const
\end{equation}
at large $t$ and the resulting fit parameters $w_0$, $E_0$ and $const$ are given in Table \ref{tab_I_zero}. All  terms in (\ref{c_bc}) are equally important near $t\simeq T/2$ since $(e^{-E_0 t}+e^{-E_0 (T-t)})_{t=T/2}/const=2e^{-E_0 T/2}/const$ is of the order one, as shown in Table \ref{tab_I_zero}. 
The effective mass computed after subtracting the resulting constant  (\ref{lambda_bc})
\begin{equation}
\label{meff_bc}
\frac{\lambda_0^{\pi\pi}(t)-const}{\lambda_0^{\pi\pi}(t+1)-const}=\frac{e^{-m_{eff}^{t+1/2}~t}+e^{-m_{eff}^{t+1/2}~(T-t)}}{e^{-m_{eff}^{t+1/2}~(t+1)}+e^{-m_{eff}^{t+1/2}~(T-t-1)}}
\end{equation}
is plotted by full symbols in Figs. \ref{fig_eigenvalues} and \ref{fig_corrected}. It has a plateau at large $t$ in contrast to the effective mass computed from (\ref{meff}), which is another indication that the ground state eigenvalue corresponds to $\pi\pi$. 

\vspace{0.2cm}

The $I=1/2$ ground state $\pi(0) K(0)$ is fitted to 
\begin{equation}
\label{lambda_bc2}
\lambda_0^{\pi K}(t)=w_0 [ e^{-E_0 t}+e^{-E_0 (T-t)}]+ A [e^{-m_{\pi} t}e^{-m_{K} (T-t)} + e^{-m_{K} t}e^{-m_{\pi} (T-t)}] 
\end{equation}
and the resulting fit parameters $E_0$, $w_0$ and $A$ are given in Table \ref{tab_I_half}. Note that $A$ is comparable to $w_0$, so all four terms are 
 of similar size near $t\simeq T/2$. 
The  $m_\pi$ and $m_K$ were fixed to the measured values in (\ref{lambda_bc2}) and we verified that  the variation of the results is negligible if $m_{\pi,K}$ are varied in the ranges given in Table \ref{tab_pseudo}. The effective mass computed after subtracting the last term  in (\ref{lambda_bc2})
\begin{equation}
\label{meff_bc2}
\frac{\lambda_0^{K\pi}(t)- A [e^{-m_{\pi} t}e^{-m_{K} (T-t)} + \{t\leftrightarrow T-t\}]}{\lambda_0^{K\pi}(t+1)- A [e^{-m_{\pi} (t+1)}e^{-m_{K} (T-t-1)} + \{t\leftrightarrow T-t\}]}=\frac{e^{-m_{eff}^{t+1/2}~t}+e^{-m_{eff}^{t+1/2}~(T-t)}}{e^{-m_{eff}^{t+1/2}~(t+1)}+e^{-m_{eff}^{t+1/2}~(T-t-1)}}
\end{equation}
 has a plateau at large $t$ (see Figs. \ref{fig_eigenvalues}, \ref{fig_corrected}),  which is another indication that ground state eigenvalue corresponds to the $\pi K$ scattering.

\section{ Variational analysis with several states contributing to a single eigenvalue}

A distinct feature of our spectrum is that there is no eigenstate corresponding to the first excited scattering  state $P_1(1)P_2(-1)$ (see Fig. \ref{fig_eigenvalues}). Another feature is that the effective mass for the ground state is not flat at intermediate $t$ and seems to contain important contributions from several physical states, not just $P_1(0)P_2(0)$ with a correction (\ref{lambda_n}). These  observations indicate that our interpolator basis (\ref{O_smear}) does not decouple the few lowest scattering states to separate eigenvalues and several of them contribute to the ground state eigenvalue. 

Let us present an example for a specific mechanism that may be responsible for 
 that. The point source couples to all the scattering states 
equally  
\begin{equation}
\label{point}
[\bar q\bar q][qq]_{point}=c_{point}\sum_{\vec k} g_{\vec k} |P_1(\vec k)P_2(-\vec k)\rangle +... ~,
\end{equation} 
up to a factor $g_{\vec k}$, which gives  a Lorentz structure of the coupling ($g_{\vec k}=1$ for non-derivative coupling and $g_{\vec k}=k_{1\mu}k_2^\mu$ for derivative coupling \cite{hooft}). For a spatially extended source  one expects additional $\vec k$-dependence $f_{{\cal O}_i}(\vec k)$  in (\ref{point})  given by the shape of smearing. 
Our interpolators (\ref{O_smear}) are not point, but are all rather narrow with an extent of a few lattice spacings. Let us explore the consequences of the approximation that all our sources behave close to point-like. In this case the few lowest scattering states couple to a given interpolator equally, but there is a different overall magnitude for each interpolator 
\begin{align}
[ \bar q_n\bar q_n][q_n q_n]&=c_{nnnn}\sum_{\vec k} g_{\vec k}~|P_1 (\vec k)P_2(-\vec k)\rangle +
a_{nnnn}|a\rangle+b_{nnnn}|b\rangle\nonumber\\ 
[ \bar q_w\bar q_w][q_w q_w]&=c_{wwww}\sum_{\vec k} g_{\vec k}~|P_1(\vec k)P_2(-\vec k)\rangle +
a_{wwww}|a\rangle+b_{wwww}|b\rangle\nonumber \\
[ \bar q_n\bar q_w][q_n q_w]&=c_{nwnw}\sum_{\vec k} g_{\vec k}~|P_1(\vec k)P_2(-\vec k)\rangle +
a_{nwnw}|a\rangle+b_{nwnw}|b\rangle\nonumber~. 
\end{align}
Here we assumed that our interpolators couple only to two  additional physical states $a$ and $b$. Given these linear combinations, one can construct the corresponding $3\times 3$ correlation matrix and it can be easily shown that its eigenvalues are 
\begin{align}
\label{lambda_var}
\lambda_0(t)&=w_0 ~\sum_{\vec k} g_{\vec k} ~ \frac{ e^{-(E^{P_1(\vec k)}+E^{P_2(\vec k)})t}}{E^{P_1(\vec k)}E^{P_2(\vec k)} }\simeq 
w_0\biggl[g_0~\frac{e^{-(m_{P1}+m_{P2}) t}}{m_{P1}m_{P2}}+6~g_1~\frac{e^{-(E^{P_1(1)}+E^{P_2(1)}) t}}{E^{P_1(1)}E^{P_2(1)} }+...\biggr]\nonumber\\
\lambda_a(t)&=w_a~e^{-E_a t}\nonumber\\
\lambda_b(t)&=w_b~e^{-E_b t}~.
\end{align} 
The physical states $a$ and $b$ get their own exponentially falling eigenvalues, while a whole tower of scattering states contributes to a single eigenvalue in this approximation. Corrections to (\ref{lambda_var}) come from additional heavy  physical states and from  slight $\vec k$-dependence of shape functions $f_{{\cal O}_i}(\vec k)$ for our sources. We note that   
 a  related variational problem has been studied in  \cite{bgr_ghosts}, where  states with non-exponential time dependence (ghosts) were found to contribute to a separate eigenvalue. 

\begin{figure}[tbh!]
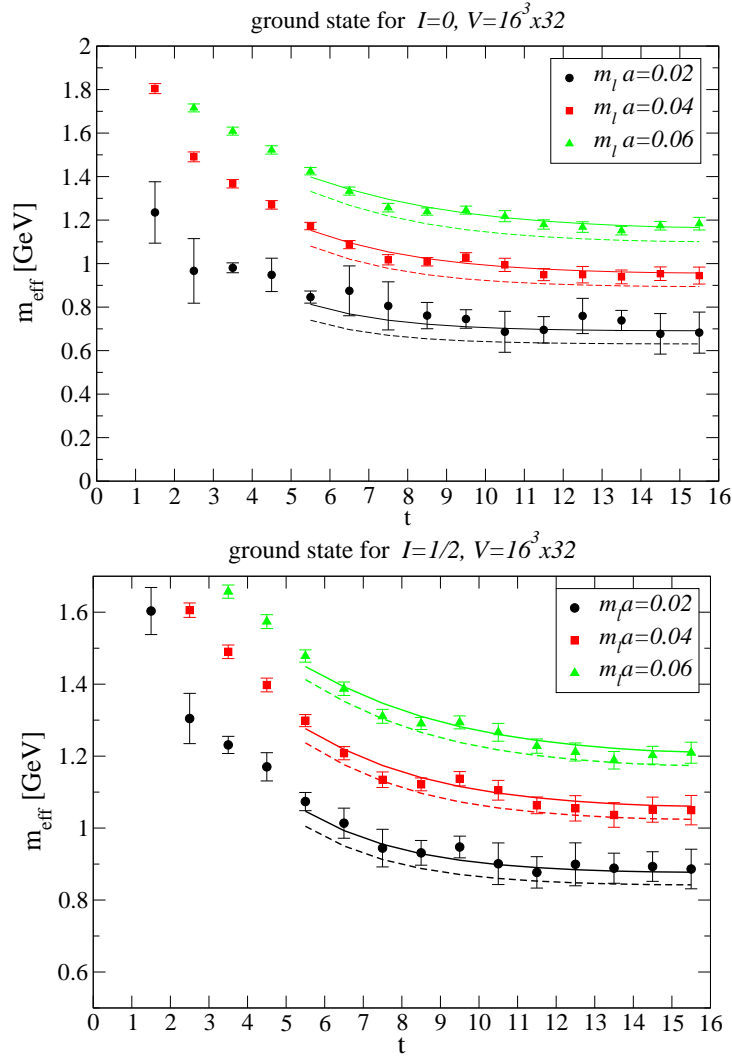

\begin{center}
\includegraphics[height=7cm,clip]{fig/eff_1_scalar_I_zero_phy_compare_16x32.eps}
\includegraphics[height=7cm,clip]{fig/eff_1_scalar_I_half_phy_compare_16x32.eps}
\end{center}
\caption{ \small Symbols present effective masses (\ref{meff_bc}) for the ground state eigenvalues in $I=0,1/2$ channels. Lines are effective masses for analytic prediction  (\ref{lambda_var}) where a tower of $\pi\pi$ ($I=0$) or $\pi K$ ($I=1/2$) scattering states contributes to $\lambda_0(t)$.  We sum  (\ref{lambda_var}) over five lowest $|\vec k|$ and verify that the contribution of higher $|\vec k|$ is negligible  in the plotted time range. We use $g_{\vec k}=1$, the measured values of $m_{\pi,K}$ and the ground state energies $m_{P_1}+m_{P_2}$ (full lines) or $m_{P_1}+m_{P_2}+dE^{tree}$ (dashed lines). }\label{fig_compare}
\end{figure}

We find that the described scenario describes our ground state eigenvalues   rather well. Fig. \ref{fig_compare}  shows that the effective masses (\ref{meff_bc},\ref{meff_bc2})  agree quite well  with the effective masses for the analytic prediction $\lambda_0(t)$  in (\ref{lambda_var}) with $g_{\vec k}=1$. We conclude that (i) a coupling $[\bar q\bar q][qq]\to P_1P_2$ is close to a non-derivative coupling with $g_{\vec k}=1$ for our interpolators and (ii) a tower of few lowest scattering states $\pi\pi,~ K\pi$ is indeed contributing to our ground state eigenvalues for $I=0,1/2$.

\newpage
\begin{table}[h]
\begin{center}
\begin{tabular}{c c c c c c c}
\hline
$m_la$ & $m_{s}a$ & $m_\pi a$ & $m_\pi$ [MeV] & $m_K a$ &  $m_K$ [MeV] & fit-range  \\ 
\hline
$0.02$ & $0.08$ &  $0.259(3)$ & $344(4)$ & $0.397(3)$ & $528(4)$ & 7-12\\
$0.04$ & $0.08$ &  $0.357(3)$ & $475(4)$ & $0.433(3)$ & $576(4)$ & 7-12\\
$0.06$ & $0.08$ &  $0.433(3)$ & $576(4)$ & $0.466(3)$ & $620(4)$ & 7-12\\
$0.08$ & $0.08$ &  $0.500(3)$ & $665(4)$ & $0.500(3)$ & $665(4)$ & 7-12\\
\hline
\end{tabular}
\end{center}
\caption{ \small The pion and kaon masses from our quenched simulation at $V=16^3\times 32$ and $a^{-1}=1.33$ GeV, obtained from the ground state eigenvalue of $3\times 3$ correlation matrix presented in \cite{bgr_mesons}.  The values at $12^3\times 24$ agree within the errors.   }\label{tab_pseudo}
\end{table}

\begin{table}[h]
\begin{center}
\begin{tabular}{c |  c c |  c c c c | c c  }
$I=0$ & $m_la$ & $V$  & $E_n a$ & $w_n$ & fit-range & ${\displaystyle\frac{\chi^2}{d.o.f.}}$ & $const$ & ${\displaystyle\frac{2e^{-E_0 T/2}}{const}}$   \\ 
\hline
\hline 
       &  $0.02$ & $16^3 32$ & $0.54(3)$ & $1.8(6) \cdot 10^{15}$ & 9-16 & $0.01$ & $6.1(16)\cdot 10^{11} $ & $1.0$ \\
{\small ground} &  $0.04$ & $16^3 32$ & $0.71(2)$ & $5.1(13) \cdot 10^{14}$ & 11-16 & $0.002$ & $5.8(13)\cdot 10^9$ & $2.0$ \\
{\small state}    &  $0.04$ & $12^3 24$ & $0.72(5)$ & $1.3(5) \cdot 10^{15}$ & 8-12 & $0.04$ & $2.6(11) \cdot 10^{11}$ & $1.7$ \\
                  &  $0.06$ & $16^3 32$ & $0.88(2)$ & $4.4(7) \cdot 10^{14}$ & 11-16 & $0.02$ & $3.1(4)\cdot 10^8$ & $2.3$ \\
       &  $0.06$ & $12^3 24$ & $0.86(3)$ & $8.8(23) \cdot 10^{14}$ & 8-12 & $0.03$ & $1.9(9) \cdot 10^{10}$ & $3.0$ \\
\hline 
                  & $0.04$ & $16^3~32$ & $1.63(5)$ & $1.1(2)\cdot 10^{13}$ & 3-8 & $0.07$ &  &  \\
{\small first}    & $0.04$ & $12^3~24$ & $1.57(9)$   & $1.1(3)\cdot 10^{13}$ & 3-7 & $0.07$ &  &  \\
{\small exc.}     & $0.06$ & $16^3~32$ & $1.70(4)$ & $9.9(13)\cdot 10^{12}$ & 3-8 & $1.5$ &  &  \\
{\small state}    & $0.06$ & $12^3~24$ & $1.69(6)$ & $1.1(2)\cdot 10^{13}$ & 3-7 & $0.1$ &  &  \\
\hline
                  & $0.04$ & $16^3~32$ & $1.88(5)$ & $4.4(7)\cdot 10^{11}$ & 3-6 & $0.1$ &  &  \\
{\small sec.}     & $0.04$ & $12^3~24$ & $1.96(8)$ & $5.6(14)\cdot 10^{11}$ & 3-6 & $0.4$ &  &  \\
{\small exc.}     & $0.06$ & $16^3~32$ & $1.97(4)$ & $4.4(5)\cdot 10^{11}$ & 3-7 & $0.2$ &  &  \\
{\small state}    & $0.06$ & $12^3~24$ & $1.97(5)$ & $4.5(8)\cdot 10^{11}$ & 3-7 & $0.7$ &  &  \\
\hline
\end{tabular}
\end{center}
\caption{ \small The energy levels $E_n$ and spectral weights $w_n$ from three eigenvalues in $I=0$ channel, for various $m_l$ and $V$. The energies in physical units are obtained by multiplying $Ea$ with $a^{-1}\simeq 1.33$ GeV. The ground state eigenvalue is fitted to (\ref{lambda_bc}), which contains additional  $const$ and its relative importance at $t=T/2$ is represented by $2e^{-E T/2}/const$. For the lowest mass $m_la=0.02$ some results are not presented as they are to noisy.}\label{tab_I_zero}
\end{table}

\begin{table}[h]
\begin{center}
\begin{tabular}{c |  c c |  c c c c | c  }
$I=1/2$ & $m_la$ & $V$  & $E_n a$ & $w_n$ & fit-range & ${\displaystyle\frac{\chi^2}{d.o.f.}}$ & $A$ \\
\hline
\hline 
       &  $0.02$ & $16^3 32$ & $0.67(3)$ & $9.2(29) \cdot 10^{14}$ & 10-16 & $0.001$ & $5.3(13)\cdot 10^{14} $  \\
{\small ground} &  $0.04$ & $16^3 32$ & $0.79(3)$ & $4.3(13) \cdot 10^{14}$ & 12-16 & $0.005$ & $2.0(5)\cdot 10^{14}$  \\
{\small state}    &  $0.04$ & $12^3 24$ & $0.78(4)$ & $1.0(3) \cdot 10^{15}$ & 8-12 & $0.04$ & $4.7(22) \cdot 10^{14}$ \\
                  &  $0.06$ & $16^3 32$ & $0.90(2)$ & $3.8(8) \cdot 10^{14}$ & 12-16 & $0.01$ & $1.3(3)\cdot 10^{14}$  \\
       &  $0.06$ & $12^3 24$ & $0.90(3)$ & $8.8(21) \cdot 10^{14}$ & 8-12 & $0.03$ & $2.8(12) \cdot 10^{14}$  \\
\hline 
                  & $0.04$ & $16^3~32$ & $1.66(4)$ & $1.0(2)\cdot 10^{13}$ & 3-8 & $0.5$ &   \\
{\small first}    & $0.04$ & $12^3~24$ & $1.64(8)$   & $1.1(3)\cdot 10^{13}$ & 3-7 & $0.1$ &    \\
{\small exc.}     & $0.06$ & $16^3~32$ & $1.72(4)$ & $9.7(13)\cdot 10^{12}$ & 3-8 & $1.9$ &    \\
{\small state}    & $0.06$ & $12^3~24$ & $1.72(5)$ & $1.2(2)\cdot 10^{13}$ & 3-7 & $0.1$ &    \\
\hline
                  & $0.04$ & $16^3~32$ & $1.93(4)$ & $4.5(6)\cdot 10^{11}$ & 3-6 & $0.1$ &   \\
{\small sec.}     & $0.04$ & $12^3~24$ & $1.97(7)$ & $5.1(11)\cdot 10^{11}$ & 3-6 & $0.4$ &    \\
{\small exc.}     & $0.06$ & $16^3~32$ & $2.00(4)$ & $4.5(5)\cdot 10^{11}$ & 3-7 & $0.2$ &    \\
{\small state}    & $0.06$ & $12^3~24$ & $1.98(5)$ & $4.5(7)\cdot 10^{11}$ & 3-7 & $0.7$ &   \\
\hline
\end{tabular}
\end{center}
\caption{ \small 
Same as Table \ref{tab_I_zero} but for $I=1/2$ channel.  The ground state eigenvalue is fitted to (\ref{lambda_bc2})  with an additional term which is proportional to the parameter $A$.  }\label{tab_I_half}
\end{table}

\begin{table}[h]
\begin{center}
\begin{tabular}{c |  c c |  c c c c   }
$I=1$ & $m_la$ & $V$  & $E_n a$ & $w_n$ & fit-range & ${\displaystyle\frac{\chi^2}{d.o.f.}}$  \\
\hline
\hline 
                   &  $0.02$ & $16^3 32$ & $0.68 - 0.82$ & &  &    \\
  {\small ground}  &  $0.04$ & $16^3 32$ & $0.78 - 0.92$ & &  &    \\
  {\small state}   &  $0.06$ & $16^3 32$ & $0.88 - 1.00$ & &  &    \\
 \hline           
                   & $0.02$ & $16^3~32$ & $1.65(4)$ & $1.1(2)\cdot 10^{13}$ & 3-6 & $0.05$    \\
                   & $0.02$ & $12^3~24$ & $1.57(10)$ & $1.1(3)\cdot 10^{13}$ & 3-6 & $0.1$    \\
                  & $0.04$ & $16^3~32$ & $1.69(4)$ & $1.0(1)\cdot 10^{13}$ & 3-8 & $1.2$   \\
{\small first}    & $0.04$ & $12^3~24$ & $1.67(6)$   & $1.1(2)\cdot 10^{13}$ & 3-7 & $0.1$   \\
{\small exc.}     & $0.06$ & $16^3~32$ & $1.73(4)$ & $9.7(12)\cdot 10^{12}$ & 3-8 & $2.2$   \\
{\small state}    & $0.06$ & $12^3~24$ & $1.73(5)$ & $1.2(2)\cdot 10^{13}$ & 3-7 & $0.2$  \\
\hline
                  & $0.02$ & $16^3~32$ & $1.94(5)$ & $4.6(8)\cdot 10^{11}$ & 3-6 & $0.1$    \\
{\small sec.}     & $0.02$ & $12^3~24$ & $1.91(8)$ & $4.4(11)\cdot 10^{11}$ & 3-6 & $0.4$     \\
 {\small exc.}   & $0.04$ & $16^3~32$ & $1.97(4)$ & $4.4(5)\cdot 10^{11}$ & 3-6 & $0.3$    \\
{\small state}     & $0.04$ & $12^3~24$ & $1.96(5)$ & $4.4(7)\cdot 10^{11}$ & 3-6 & $0.8$     \\
     & $0.06$ & $16^3~32$ & $2.01(4)$ & $4.4(5)\cdot 10^{11}$ & 3-7 & $0.3$    \\
    & $0.06$ & $12^3~24$ & $1.99(5)$ & $4.3(6)\cdot 10^{11}$ & 3-7 & $0.9$    \\
\hline
\end{tabular}
\end{center}
\caption{ \small 
Same as Table \ref{tab_I_zero} but for $I=1$ channel. We are unable to perform a two-state ($K\bar K,~\pi\eta_{ss}$) fit for the ground state eigenvalue, so we provide only informative energy ranges given by the naive cosh-type effective mass $m_{eff}^{9.5}-m_{eff}^{13.5}$.  }\label{tab_I_one}
\end{table}

\end{document}